\documentstyle[prb,aps,epsf]{revtex}
\begin{document}
\tighten
\twocolumn
\title { Morphology transitions in three-dimensional domain growth \\
 with Gaussian random fields}
\author{Belita Koiller }
\address{Instituto de F\'\i sica, Universidade Federal do Rio de Janeiro, 
Cx.P. 68.528, 21945-970, RJ, Brazil}
\author{Mark O. Robbins } 
\address{ Department of Physics and Astronomy, Johns Hopkins University, 
Baltimore, MD 21218}
\date{March 31, 2000}
\maketitle
\begin{abstract}
We study the morphology of magnetic domain growth in disordered three
dimensional magnets.
The disordered magnetic material is described within the random-field 
Ising model with a Gaussian distribution of local fields with width
$\Delta$.
Growth is driven by a uniform applied magnetic field, whose value is kept 
equal to the critical value $H_c(\Delta)$ for the onset of steady  motion.
Two growth regimes are clearly identified.
For low $\Delta$ the growing domain is 
compact, with a self-affine external interface.
For large $\Delta$ a 
self-similar percolation-like morphology is obtained. 
A multi-critical point at $(\Delta_c$, $H_c(\Delta_c))$ separates
the two types of growth.
We extract the critical exponents near $\Delta_c$ using finite-size
scaling of different
morphological attributes of the external domain interface.
We conjecture that the critical disorder width also corresponds to a 
maximum in $H_c(\Delta)$.
\end{abstract}
%\pacs{75.60.Ch,68.35.Rh,75.60.Ej}

\section{Introduction}
\label{sec:1}
Many physical processes involve motion of the interface between two domains
through a disordered media.  Classic examples are magnetic domain growth,
fluid invasion of porous media and fluid segregation in
gels.\cite{familybook,stanleybook}
Previous studies of such systems
have shown that changing the strength of disorder
leads to critical transitions between different growth
morphologies.\cite{cieplak,martys,ji2d,ji3d,kjr1,kjr2,sethna1,sethna2,stokes}
Most theoretical work on the subject has been done on
models with an underlying crystalline lattice of spins or pores.
The crystalline anisotropy may lead to a faceted growth regime in
the low disorder limit.
As disorder increases, there may be a transition to
compact growth with a self-affine interface.
In the high disorder limit, growing domains are self-similar fractals,
characteristic of percolation.

Studies of driven interfaces in
the zero-temperature random-field (and random-bond) Ising model (RFIM)
indicate that the dimensionality, coordination number, and distribution
of random fields are all important in determining the sequence of
morphological transitions.\cite{ji2d,ji3d,kjr1}
In two dimensions (2D), there is a transition from self-similar to faceted
growth at a critical value of disorder, {\em if} the
the distribution of random fields is bounded.\cite{ji2d}
The critical behavior is not universal, and can be related to the
analytic form of the tails in the distribution.\cite{kjr1}
For an unbounded Gaussian distribution, percolative growth occurs for
{\em any} finite strength of disorder.
In the three dimensional (3D) RFIM with a bounded distribution of
disorder all three types of growth morphology,
faceted, self-affine, and percolative,
are found as the strength of disorder increases.\cite{ji3d}
Analysis of growth probabilities suggests that the transition
from percolative to self-affine growth might be universal,
but that the self-affine to faceted transition is non-universal and
could be suppressed by an unbounded distribution of fields.

The suggestion that faceted growth may be eliminated by unbounded
distributions of disorder in 3D is supported by
renormalization group studies of {\em equilibrium} interface
conformations.\cite{nattermanprl}
Calculations for a Gaussian distribution of random fields show
that faceted interfaces are suppressed by any amount of
disorder in dimensions $d \leq 3$.
Porous media and some magnetic systems do not have any underlying
crystalline structure, and it is only introduced in the models for
computational convenience.
Thus suppression of faceted growth by unbounded distributions of
disorder leads lattice models to provide a more accurate description
of real systems. 

In this paper we explore the effect of unbounded Gaussian disorder
on growth morphology transitions in the 3D RFIM at zero temperature.
We find that this unbounded distribution does indeed eliminate the faceted
growth regime.
There is a single transition from self-affine growth at
low disorder to percolative growth at high disorder.
For each value of the disorder we find the critical field $H_c$ needed to
initiate steady growth.
The critical behavior at the onset of growth is analyzed, and the critical
exponents are consistent with previous results for bounded distributions of
random fields.

The transition between self-similar and self-affine growth is shown to be a
multi-critical point.
We identify lengths that diverge as the strength of disorder is varied
along the line of critical fields, and evaluate critical exponents.
Our analysis reveals problems with previous work on bounded distributions of
random fields.
These studies identified a ``fingerwidth'' with the correlation length that
diverges at the multi-critical point.\cite{ji3d}
However, our work with larger systems shows that this fingerwidth does not
diverge,
and its saturation leads to errors in the determination of critical properties.
New work will be needed to determine whether the self-affine to percolative
transitions
for bounded and unbounded distributions are in the same universality class.
However, we can compare our results to studies of magnetic hysteresis in the 3D
RFIM with Gaussian random fields by
Perkovi\'c {\em et al.}.\cite{sethna1,sethna2}
These studies use a different growth algorithm, and
we find that this leads to changes in the critical disorder and critical
exponents.

The paper is organized as follows.
In Sec.~\ref{sec:2} we describe the growth model. 
Studies of the critical field and critical exponents at the onset of motion
are presented in Sec. \ref{subsec:a}.
The transition in growth morphology is examined in Sec. \ref{subsec:b}, and
Sec.~\ref{sec:5} contains our general summary and conclusions.

\section{Growth model and algorithm}
\label{sec:2}
The energy of a system of Ising spins ($s_i = \pm 1$)
at the sites $i$ of a simple cubic lattice is written as 
\begin{equation}
{\cal H}= - \sum _{<i,j>} s_i s_j -\sum _i (\eta_i + H) s_i ~.
\label{H}
\end{equation}
The first term on the right-hand side
of Eq. (\ref{H}) represents the ferromagnetic coupling 
between nearest neighbor spins, with the exchange coupling
taken as the energy unit.
The second term gives the interaction of each spin with the uniform external 
magnetic field $H$ and with the random local field $\eta_i$.  
The local fields are uncorrelated, following a Gaussian distribution 
function whose width $\Delta$ quantifies the degree of disorder:
\begin{equation}
 P(\eta)=(2\pi\Delta^2)^{-1/2} e^{-\eta^2/(2\Delta^2)}~.
\label{gauss}
\end{equation}

As in previous studies,\cite{ji3d} the simulation cell is a 
cube of side $L$, with the lattice constant taken as the unit of length.
Periodic boundary conditions are imposed along the $x$ and $y$ directions.
All spins are initially anti-parallel to the external field, i.e. $s_i = -1$, 
except those in the bottom layer, $z=1$. 
This layer constitutes the ``seed'' for growth of the +1 domain, which is 
driven by the external field.
The orientation of this seed plane does not affect critical behavior 
in the self-affine and self-similar growth regimes of interest
here,\cite{kjr1,martysthesis,barabasi}
but does affect the faceted growth seen for weak, bounded distributions
of disorder.\cite{kjr1,nowak1,nowak2}

Growth proceeds through single spin flips at zero temperature. 
Whenever a spin flip from $s_i =-1$  to  $s_i=+1$  lowers the total energy, 
it is implemented. 
However, {\em only spins on the growing interface are allowed to flip}. 
This restriction is motivated by the fluid invasion
problem,\cite{ji2d,ji3d,kjr1}
and differentiates our dynamics from the model considered 
by Perkovi\'c {\em et al.}\cite{sethna1,sethna2} in studies
of hysteresis in magnetic systems.

We have developed a new memory-efficient algorithm that allows us to consider
cells with $L$ as large as 1152 in the present study.
The simulation cell is subdivided into smaller cubic cells of side $s$.
The $s^3$ spins of a sub-cell remain active in memory only while at least 
one of them is a ``flippable'' interface spin.
Growth occurs at fixed driving field $H$.
This allows the local random field to be encoded into a single byte that gives
the minimum number of additional neighbors needed to flip the spin at that site.
This number is decreased by one every time the spin acquires a new $+1$ 
neighbor. 
When enough neighbors are present, the spin is flipped. 
For reasons discussed in Sec.~\ref{subsec:b}, we flipped 
all spins that were completely surrounded by $+1$ neighbors, regardless 
of their local field.
Such spins can not affect growth at any other site, and being able to
remove the cells containing them from memory allowed us 
to treat larger systems. 
Preliminary runs where these spins were not flipped gave equivalent results.

At each $L$ and $\Delta$, domains
were grown in an ensemble of samples with different configurations of
$\{\eta_i\}$.
In each sample,
growth was stopped when all spins on the interface were stable, or when 
the interface first reached the top of the system ($z=L$).
The morphology of the domain was then analyzed as described below.

\section{Results}
\label{sec:3}
\subsection{The critical field}
\label{subsec:a}

\begin{figure}[bt]
%\setlength{\unitlength}{1mm}
%\begin{picture}(100,150)(0,0)
\begin{picture}(150,250)(0,0)
\put(-45,-35){\epsfxsize=10cm\epsfbox{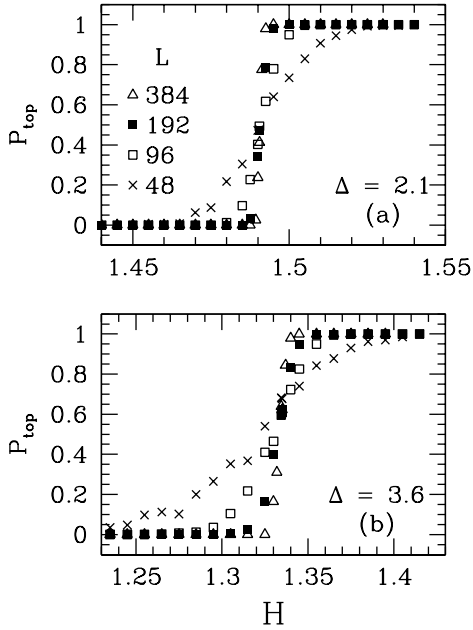}}
%\put(-10,-15){\epsfxsize=16cm\epsfbox{figuram1.eps}}
\end{picture}
\caption{Probability $P_{top}$ that the domain grows to the top of the system
as a function of $H$ at the indicated system sizes $L$ for
(a) $\Delta = 2.1$ (self-affine) and (b) $\Delta = 3.6$ (self-similar).
Four hundred different systems were used for each $L$ and $\Delta$,
so the statistical uncertainty is less than $\pm 0.025$ (symbol size).
}
\label{fig:field} 
\end{figure}

The external field $H$ provides a driving force that causes the
domain of $+1$ spins to grow.
If $H$ is too small, the domain wall will remain pinned near the
bottom of the system, while large $H$ will cause all spins to flip.
Figure \ref{fig:field} shows the probability, $P_{top}$,
for a domain to grow to
the top of the cell as a function of $H$ at several
different system sizes.
Results for two values of $\Delta$ corresponding to self-affine
($\Delta = 2.1$) and self-similar ($\Delta = 3.6$) growth are shown.
For each $\Delta$,
the range of $H$ over which the probability rises from 0 to 1 becomes
narrower as $L$ increases.
All of the curves intersect at a critical field $H_c(\Delta)$ and
probability $P_c(\Delta)$.
If $H >H_c$, the probability approaches unity in the thermodynamic limit
$(L \rightarrow \infty)$,
while the probability vanishes in this limit if $H < H_c$.

Previous studies with a uniform distribution of random
fields\cite{cieplak,martys,ji2d,ji3d,kjr1,kjr2,nolle,nollethesis}
have shown that there is a diverging correlation length
$\xi_H \sim |H-H_c|^{-\nu_H}$ as $|H-H_c| \rightarrow 0$.
The critical exponent in the self-similar growth regime\cite{ji3d}
is consistent with the 3D percolation exponent,\cite{adler90}
$\nu_H=0.88 \pm 0.02$, while $\nu_H=0.75\pm 0.02$ for
self-affine growth.\cite{ji3d,nollethesis}

\begin{figure}[bt]
%\setlength{\unitlength}{1mm}
%\begin{picture}(150,150)(0,0)
\begin{picture}(150,250)(0,0)
\put(-43,-33){\epsfxsize=10cm\epsfbox{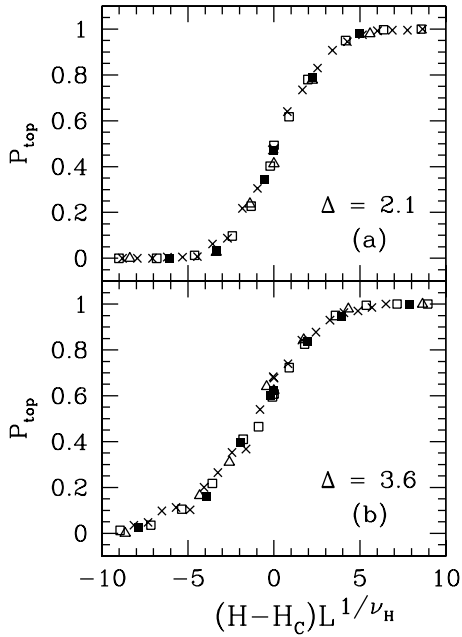}}
%\put(-10,-15){\epsfxsize=15cm\epsfbox{figuram2.eps}}
\end{picture}
\caption{Probability that the domain grows to the top of the system
as a function of $(H-H_c)L^{1/\nu_H}$ for
(a) $\Delta = 2.1$ and (b) $\Delta = 3.6$.
In (a) $H_c = 1.4905$ and $\nu_H = 0.75$.
For (b) $H_c =1.335$ and $\nu_H = 0.88$.
The symbols for each $L$ are the same as in Fig. \ref{fig:field}.
Four hundred different systems were used for each $L$ and $\Delta$,
so the statistical uncertainty is less than $\pm 0.025$ (symbol size).
}
\label{fig:colhc} 
\end{figure}

Near $H_c$, the probability for the domain to span the system should
only depend on the ratio of system size to $\xi_H$.
This suggests that when probability is plotted against $(H-H_c) L^{1/\nu_H}$
the results for all system sizes should collapse onto a universal curve.
Fig. \ref{fig:colhc} verifies that the data from Fig. \ref{fig:field}
are consistent with this ansatz, and with the values of $\nu_H$ that
were found for uniform distributions of random fields.


Examination of the spanning probability for other values of $\Delta$
shows that $P_c(\Delta)$ is always near 2/3 for self-similar growth
and decreases with $\Delta$ in the self-affine regime.\cite{foot1}
To determine $H_c(\Delta)$ we worked with the largest accessible
system size ($L=$768 or 1152) and found the value of $H$ that gave a
spanning probability between 0.5 and 0.7.
This determines $H_c$ with an accuracy of about 0.0001.
We found that even larger errors in the growth field did not change the
morphology of spanning domains that is analyzed in the next section.

\begin{figure}
%\setlength{\unitlength}{1mm}
%\begin{picture}(80,90)(0,0)
\begin{picture}(160,180)(0,0)
\put(-70,-99){\epsfxsize=16cm\epsfbox{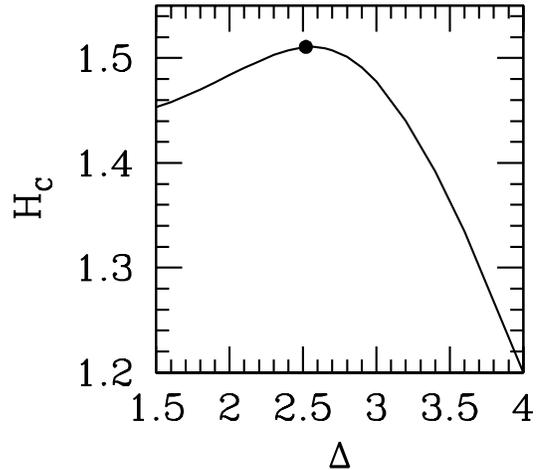}}
%\put(-20,-15){\epsfxsize=25cm\epsfbox{figura0.eps}}
\end{picture}
\caption{Variation of the critical field $H_c$ with Gaussian disorder $\Delta$.
A filled circle indicates the value of $\Delta_c=2.52\pm0.03$ determined below.
Self-affine growth occurs for $\Delta<\Delta_c$ and self-similar growth
occurs at larger $\Delta$. }
\label{fig-Hc} 
\end{figure}

In  Figure \ref{fig-Hc} we show the variation of $H_c$ 
with disorder for the range of interest in the present study.
At small $\Delta$ the value of $H_c$ increases monotonically.
There is a maximum near $\Delta=2.5$, and then $H_c$ drops monotonically,
becoming negative for $\Delta \gtrsim 7$.

\subsection{Transition in Growth Morphology}
\label{subsec:b}

As expected, the faceted growth regime seen for bounded distributions
of random fields\cite{ji3d} was suppressed by Gaussian disorder.
The only transition that we observed was from self-similar growth at
high disorder to self-affine growth at low disorder.

Self-similar growth is isotropic, while self-affine growth has
a well-defined direction at long-length scales.
Previous work on other models\cite{cieplak,martys,ji2d,ji3d,kjr1,kjr2}
shows that the transition between these two
growth regimes is a multi-critical point at some $H_c$ and $\Delta_c$.
An order parameter can be defined in analogy to equilibrium
magnetic transitions as the average of the unit vector normal
to the interface.\cite{martys,martysthesis}
For $\Delta < \Delta_c$ this average is finite,
while for $\Delta > \Delta_c$ the order parameter vanishes.
A correlation length, $\xi$, that diverges at $\Delta_c$ can also be
defined.
As $\Delta$ approaches $\Delta_c$ from above (self-similar regime),
longer and longer segments of the domain wall advance in the same
direction.
In the self-affine regime,
deviations from the mean direction occur over longer and longer length
scales as $\Delta$ increases to $\Delta_c$.
In the following subsections we examine the morphology of domains
and use finite-size scaling to determine $\Delta_c$ and the exponent
$\nu$ that describes the divergence of $\xi$
as $\Delta \rightarrow \Delta_c$.

\subsubsection{Fingerwidths}

\begin{figure}
\begin{picture}(150,250)(0,0)
%\begin{picture}(80,90)(0,0)
\put(-45,-13){\epsfxsize=10cm\epsfbox{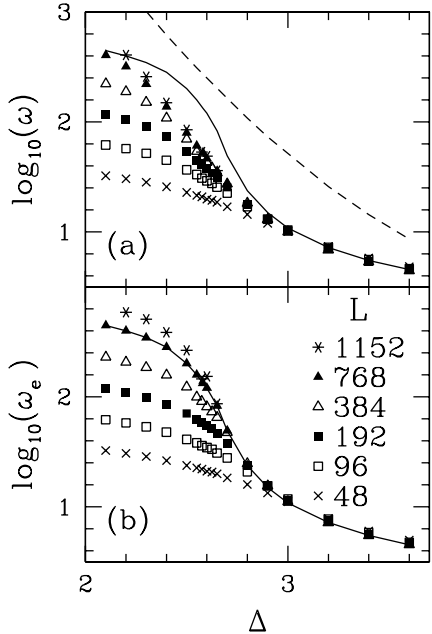}}
%\put(0,5){\epsfxsize=16cm\epsfbox{figura1.eps}}
\end{picture}
\caption{(a) Fingerwidths $w$ calculated as a function of disorder $\Delta$ 
for the simulation cell sizes indicated in (b). 
The dashed line gives an upper bound, obtained from the statistical 
probability of singlets or pairs of unflippable $s=-1$ spins.
The solid line shows the external fingerwidth $w_e$ for $L=768$.
Panel (b) shows $w_e$ vs. $\Delta$ for all $L$.
In both panels fingerwidths were calculated in the plane normal to
the nominal growth direction, $z$.
}
\label{fig-w} 
\end{figure}

Previous experimental\cite{stokes} and
theoretical\cite{cieplak,martys,ji2d,ji3d,kjr1,kjr2}
studies have used a simple measure of
the range of correlations.
A fingerwidth $w$ was calculated by examining lines of adjacent
nearest-neighbor spins,
and averaging the length of contiguous segments of $+1$ spins
(or fluid-invaded regions\cite{cieplak,martys,kjr2,stokes}).
Results obtained from our simulations
are given in Figure \ref{fig-w}(a).
As in previous work, $w$ is independent of $L$ at high disorder
and proportional to $L$ at low disorder.
If one assumes that $w$ diverges at $\Delta_c$ in the thermodynamic limit,
$w\propto (\Delta - \Delta_c)^{-\nu}$, then one can determine $\Delta_c$
and $\nu$ from finite-size scaling collapses of the fingerwidth data.
While this assumption was used in previous studies, our results with
larger system sizes indicate that it is not justified.
Finite-size scaling collapses become worse and worse as the range of $L$ 
increases, and give estimates for $\Delta_c$ that are clearly in the
self-affine regime.

The reason that $w$ does not diverge at $\Delta_c$ is quite simple.
Even in the self-affine regime there
are small clusters of unflippable spins (Fig. \ref{fig-cross}) due to
the tails in the distribution of $\eta_i$. 
These clusters are left behind by the advancing interface, and do
not affect the morphology at long-length scales.
However they do lead to a finite value of $w$.
The limiting value of $w$ can be estimated from the probability
for single isolated spins, pairs of spins, etc..
The upper bound for $w$ from single spins and pairs is indicated
by a dashed line in Fig. \ref{fig-w}(a).
This upper bound clearly inhibits divergence of the fingerwidth with $L$
at any finite $\Delta$.
It is only about 300 at the value of $\Delta_c \approx 2.5$ determined below,
and lower bounds would be obtained by considering larger clusters.

We attempted to define improved fingerwidths by
eliminating all isolated spins, and
then all pairs of isolated spins before determining the fingerwidth.
However, this procedure was inefficient and did not converge rapidly.
In the following we focus entirely on the morphology of the external
interface, and thus eliminate surrounded unflipped regions of all sizes.
The scaling behavior in the self-affine regime is only associated with
the external interface:
Once the surrounded regions are left behind, they become irrelevant.
In the self-similar regime we know that the external interface of
a percolation cluster has the same fractal dimension as the entire cluster.
Thus the external interface should give nearly the same fingerwidth as
the cluster.

The solid line in Figure \ref{fig-w}(a) shows the fingerwidth 
calculated from the external interface, $w_e$, as a function of $\Delta$
at $L=768$.
As expected, the unflipped regions have no effect on fingerwidth
for $\Delta \gtrsim 3$, where the interface forms a fractal percolation
pattern with narrow fingers.
However as $\Delta$ decreases towards $\Delta_c \approx 2.5$,
the solid line rises sharply above the other data points.
Figure \ref{fig-w}(b) shows how the external fingerwidth 
changes with $L$.
Finite-size scaling collapses of $w_e$ are discussed below.

\subsubsection{Interface roughness }

Self-affine interfaces are characterized by the scaling properties
of the interface roughness.\cite{familybook,stanleybook}
Due to our periodic boundary conditions, the average direction of
the external interface is normal to $z$.
Its position is given by a height $h(x,y)$ that may be multivalued.
The roughness $\rho(\ell)$ over a square region of side $\ell$
in the $x-y$ plane can be
quantified by the root-mean-squared (rms) variation in $h$
\begin{equation}
\rho(\ell) = \sqrt {\langle(h-\langle h \rangle )^2 \rangle _\ell}~,
\label{rugos}
\end{equation}
where $\langle h \rangle$ is the average over a given square and
$\langle \rangle_\ell$ indicates an average over all square regions of 
side $\ell$.
For a self-affine interface,\cite{familybook,stanleybook}
$\rho(\ell) \sim \ell^\alpha$ at large
$\ell$.
The roughness exponent $\alpha<1$ characterizes the degree of anisotropy,
and would be unity for a self-similar fractal.

\begin{figure}
%\setlength{\unitlength}{1mm}
%\begin{picture}(140,160)(0,0)
%\put(0,5){\epsfxsize=13cm\epsfbox{figuran.eps}}
\begin{picture}(180,220)(0,0)
\put(0,10){\epsfxsize=7.5cm\epsfbox{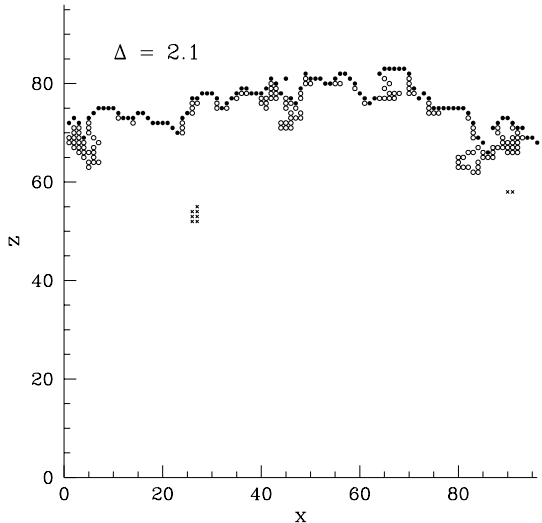}}
\end{picture}
\caption{Cross-section through a system of size $L=96$ for $\Delta = 2.1$.
The circles (open and solid) show all spins on the external interface.
Solid circles indicate the single-valued interface
obtained by taking the highest spin at each point in the (x,y) plane.
Overhangs are evident in the regions where the solid and open circles
separate.
These form as the interface grows around unflippable clusters (crosses).
As discussed in the text, unflippable clusters prevent the total fingerwidth
$w$ from diverging.
However, the interface in this figure is clearly self-affine, and the
largest unflippable cluster is much smaller than the system size.
}
\label{fig-cross}
\end{figure}

Figure \ref{fig-cross} shows a cross-section through a system of size
$L=96$ for $\Delta=2.1$, which is well into the self-affine regime.
The external interface (circles) is a multi-valued function.
In some regions there are overhangs where the interface extends over
itself.
These overhangs are necessary if the interface is to grow around small
clusters of unflippable spins (crosses).
The size of overhangs and of clusters of unflipped spins increases
as $\Delta$ rises to $\Delta_c$.

\begin{figure}
\begin{picture}(150,210)(0,0)
%\put(0,5){\epsfxsize=16cm\epsfbox{figura2.eps}}
\put(-39,-46){\epsfxsize=10cm\epsfbox{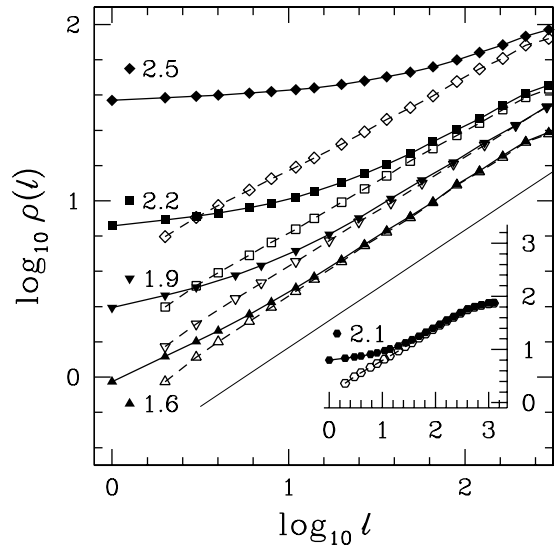}}
\end{picture}
\caption{External interface roughness at the indicated 
values of disorder $\Delta$ for $L=768$.
The full lines and symbols correspond to the total interface $\rho$,
while the dashed lines and open symbols refer to the
single-valued top interface $\rho_t$.
The straight solid line below the curves shows the slope corresponding
to $\alpha = {2\over 3}$. 
The inset at the lower right shows that $\rho$ and $\rho_t$ converge
for $\Delta = 2.1$ when a larger system size is used ($L=1260$). 
There is some rounding of the curves at large $\ell/L$ due to finite-size
effects.
}
\label{fig-rho}
\end{figure}

Previous work\cite{nolle} shows that overhangs can change
the scaling of $\rho(\ell)$ at small $\ell$.
One way of highlighting their effect is to compare $\rho(\ell)$ with
the roughness $\rho_t(\ell)$ of the single-valued interface, $h_t(x,y)$,
obtained by taking the top (highest) point on the external interface at
each $\ell$ (closed circles in Fig. \ref{fig-cross}).
Figure \ref{fig-rho} shows log-log plots of both quantities vs. $\ell$
at the indicated values of $\Delta$.
For $\Delta \leq 2.2$, results for the external (solid symbols) and
single-valued (open symbols) interfaces converge at large $\ell$.
The slope of the curves in the converged region is consistent
with the roughness exponent $\alpha = 2/3$ that is predicted from scaling
arguments\cite{ma,narayan2,natterman} and observed in previous
simulations.\cite{ji3d,nollethesis}
At small $\ell$ the dashed and full lines separate due to overhangs.
The value of $\rho_t$ goes to zero at $\ell=1$, while the
value of $\rho$ goes to the rms variation in height above a single
point in the $(x,y)$ plane.
The growing separation between dashed and full lines as $\Delta$
increases towards $\Delta_c = 2.52$
shows that the size of the overhangs increases.
For $\Delta=2.5$ there is no convergence of the lines even at the 
largest $\ell$ and $L$ we could study.
The overhang size is one measure of a diverging length as $\Delta$
approaches $\Delta_c$ from below, and is analyzed in following sections.

\subsubsection{Determining $\Delta_c$}

The data in Figures \ref{fig-w} and \ref{fig-rho} give clear evidence
of a morphological transition, and rough bounds on the value of
$\Delta_c$.
The saturation of the external fingerwidth with
increasing $L$ for $\Delta \gtrsim 2.65$
gives an upper bound for $\Delta_c$
(Fig. \ref{fig-w}(b)),
while the merging of $\rho$ and $\rho_t$ at large $\ell$ for
$\Delta \leq 2.2$ gives a lower bound (Fig. \ref{fig-rho}).
In this subsection we investigate other morphological attributes of
the external interface that provide more accurate bounds for
$\Delta_c$. 
In the next subsection we determine the exponent $\nu$ through finite size
scaling analysis of these quantities.

\begin{figure}
\setlength{\unitlength}{1mm}
%\begin{picture}(160,180)(0,0)
%\put(0,5){\epsfxsize=16cm\epsfbox{figura3.eps}}
\begin{picture}(80,90)(0,0)
\put(-8,-10){\epsfxsize=10cm\epsfbox{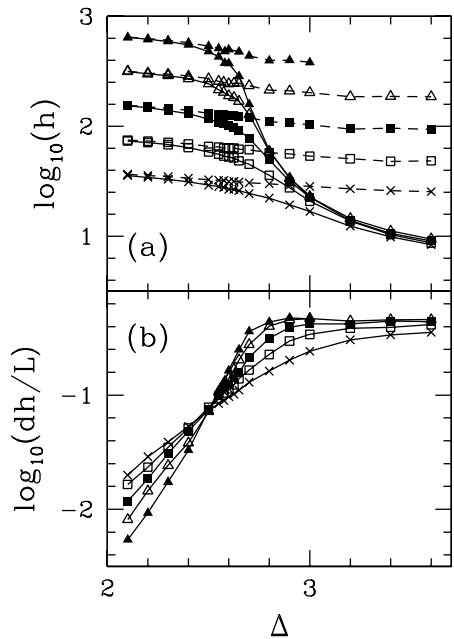}}
\end{picture}
\caption{(a) Average heights of the single-valued interfaces
$\langle h_t \rangle$ (dashed lines) and $\langle h_b \rangle$
(solid lines) as a function of $\Delta$
for system sizes ranging from $L=48$  (crosses) to $L=768$ (solid triangles), 
following the symbol definitions in Figure \ref{fig-w}.
(b) Average height difference $dh \equiv \langle h_t -h_b \rangle$,
normalized by system size, as a function of disorder. 
The crossing point where curves for all $L$ intersect
gives an estimate for $\Delta_c$.
}
\label{fig-h}
\end{figure}

A single-valued interface, $h_b(x,y)$, can also be defined by
taking the {\em bottom} (lowest) 
value of the external interface $h$ for each $(x,y)$. 
Figure \ref{fig-h}(a) shows the average heights of the top and bottom
interfaces,
$\langle h_t \rangle$ and  $\langle h_b \rangle$, 
as a function of $\Delta$ at various system sizes.
In the self-similar regime, the fractal external interface extends throughout
the entire height of the cell.
The value of $\langle h_t \rangle $ is a constant fraction of the system
size and $\langle h_b \rangle $ is of order of the fingerwidth.
The two averages converge in the self-affine regime, where the difference
between them,
$dh \equiv \langle h_t \rangle - \langle h_b \rangle $, 
is a measure of the height and abundance of overhangs.

In Figure \ref{fig-h}(b) we plot $dh /L $ vs. $\Delta$. 
As implied by the above discussion, this ratio vanishes at small $\Delta$
and rises to a constant fraction at large $\Delta$.
The increase becomes sharper with increasing $L$,
and there is a clear crossing of all curves at $\Delta \approx 2.5$.
This means that for $\Delta$ below the crossing point $dh / L$
decreases with $L$,  while above the crossing point $dh / L$
increases with system size.
We conclude that the crossing point must coincide with $\Delta_c$.

\begin{figure}
\setlength{\unitlength}{1mm}
%\begin{picture}(140,160)(0,0)
%\put(0,5){\epsfxsize=14cm\epsfbox{figura7.eps}}
\begin{picture}(70,80)(0,0)
\put(-13,-15){\epsfxsize=10cm\epsfbox{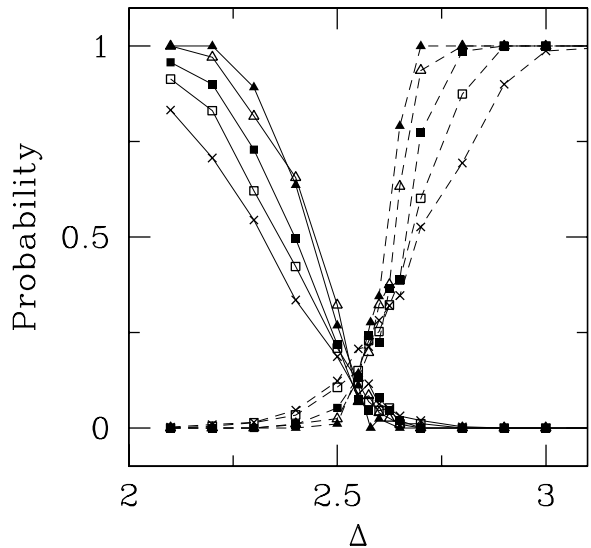}}
\end{picture}
\caption{Probabilities related to the global minimum  $h_-$ of the full
external interface.
The data points connected by solid lines give the probabilities that
$h_-\ge L/3$, while the data points connected by dashed lines give the
probability that at least one point on the external interface remains
at the initial seed plane.
System sizes range from $L=48$  (crosses) to $L=768$ (solid triangles),
following the symbol definitions in Figure \ref{fig-w}.
As $L$ increases, the probabilities show a sharper change from zero to one. 
Quantities were averaged over an ensemble of at least 100 samples
at each value of $\Delta$ and $L$.
}
\label{fig-ratio}
\end{figure}

We have examined a variety of other quantities to confirm that all give
consistent values of $\Delta_c$ and to minimize the error bars.
Figure \ref{fig-ratio} shows results for two probabilities that
are related to the global minimum, $h_-$, of each external interface.
Data points connected by solid lines give the probability that $h_-/L$
is greater than $1/3$.
This probability is unity in the self-affine limit and drops to zero
in the self-similar regime where the fractal external interface extends
all the way to the bottom of the system.
The data points connected by dashed lines in  Figure \ref{fig-ratio} give
the probability that $h_-$ remains at the height of the initial seed plane.
This probability is unity in the self-similar regime and drops to zero in
the self-affine regime.
Both probabilities exhibit sharper transitions
from one to zero as $L$ increases,
and should become step-functions at $\Delta_c$ in the limit
$L \rightarrow \infty$.
Crossing points for the two probabilities in Fig. \ref{fig-ratio},
the interface width in Fig. \ref{fig-h}(b),
and all other quantities that
we examined are consistent with $\Delta_c = 2.52 \pm 0.03$.

It is interesting to note that Fig. \ref{fig-Hc} shows a maximum in
$H_c(\Delta)$ at $\Delta_c$. 
This is a reasonable result, given the difference in growth mechanisms for
self-affine and self-similar regimes.
In the self-affine regime, the interface must advance across the entire
width of the system.
Thus $H_c$ is sensitive to the regions that 
are hardest to flip, and rises with $\Delta$.
In the self-similar regime, the interface follows the path of least resistance.
Since the number of spins that must be flipped (the percolation probability)
is less than $1/2$, increasing $\Delta$ makes it easier to flip enough
spins to span the system, and decreases $H_c$.
The rate of decrease in $H_c$
can be calculated exactly from the percolation probability
in the large $\Delta$ limit, where spins are decorrelated.\cite{ji3d,kjr1}
The asymptotic slope is:
$d H_c/d \Delta = - 0.4907 $.

\subsubsection{Finite-size scaling determination of $\nu$}
\label{sec:4}

As in Sec. \ref{subsec:a}, we use finite-size scaling to determine
the exponent $\nu$ that describes the diverging correlation length $\xi$
at $\Delta_c$.
The deviation from $\Delta_c$ is measured by
$\delta \equiv  (\Delta - \Delta_c)/\Delta$.
We assume that $\xi \sim \delta^{-\nu}$, and that
close to the critical disorder
the only relevant lengths are $\xi$ and the system size $L$.
Then dimensionless quantities like those shown in Figs.  \ref{fig-h}(b) and
\ref{fig-ratio}
can only depend on $L/\xi$, or equivalently $L^{1/\nu} \delta$.
When plotted against $L^{1/\nu}\delta$,
results for all system sizes
should collapse onto a universal scaling function.

\begin{figure}
\setlength{\unitlength}{1mm}
%\begin{picture}(160,180)(0,0)
%\put(0,5){\epsfxsize=16cm\epsfbox{figura4.eps}}
\begin{picture}(70,80)(0,0)
\put(-12,-15){\epsfxsize=10cm\epsfbox{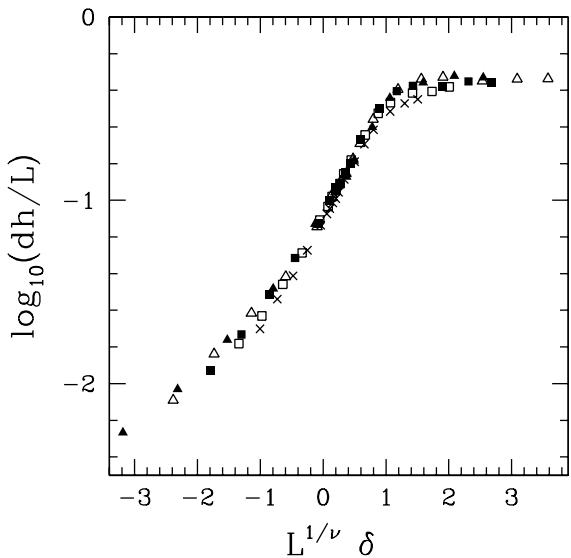}}
\end{picture}
\caption{Finite size scaling collapse of the data in Figure \ref{fig-h}
(b) for $\Delta_c=2.52$ and $\nu = 2.4$.
The symbols used for each $L$ are the same as in Figure \ref{fig-w}. 
}
\label{fig-fssh}
\end{figure}

Figure \ref{fig-fssh} shows a scaling collapse for the data
of Fig. \ref{fig-h}(b).
The scaled interface widths $dh / L$
for system sizes $L=48$, 96, 192, 384 and 768
collapse well onto a universal curve near $L^{1/\nu} \delta =0$.  
As $|L^{1/\nu} \delta|$ increases, the curves for small $L$ begin to deviate
from the others.
These deviations reflect corrections to scaling.
They appear first at small $L$ because these data points are for larger
values of $\delta$ than their counterparts at large $L$.

Since the magnitude of corrections to scaling is not known,
there is some uncertainty in determining the values of $\Delta_c$ and
$\nu$.
We found acceptable collapses for $dh/L$ with $\Delta_c=2.52\pm0.03$ and
$\nu = 2.5 \pm 0.3$.
Scaling collapses of other quantities related to the interface width,
including $\rho(1)$, the ratio of the lowest and highest points
on the entire interface, and the probabilities shown in Figure \ref{fig-ratio},
all gave consistent ranges of $\Delta_c$ and $\nu$.
We also considered the scaling variable
$\delta ^ \prime \equiv (\Delta - \Delta_c)/\Delta_c$,
which gives different, and often larger,\cite{sethna2,nolle}
corrections to scaling.
This led to a narrower region of scaling, but the same range of values
for $\nu$ and $\Delta_c$.

The scaling behavior of the external fingerwidth
is more complicated, because the distribution of fingerwidths becomes
bimodal in the self-affine regime.
Most of the fingerwidths are essentially equal to the system size $L$.
However, there is a significant fraction of very small fingerwidths
from the region of width $dh$ near the top of the interface
(see Fig. \ref{fig-cross}).
These make a disproportionate contribution to $w_e$ that does not
obey the scaling ansatz.

\begin{figure}
\setlength{\unitlength}{1mm}
%\begin{picture}(160,180)(0,0)
%\put(0,5){\epsfxsize=16cm\epsfbox{figura5.eps}}
\begin{picture}(70,75)(0,0)
\put(-12,-15){\epsfxsize=10cm\epsfbox{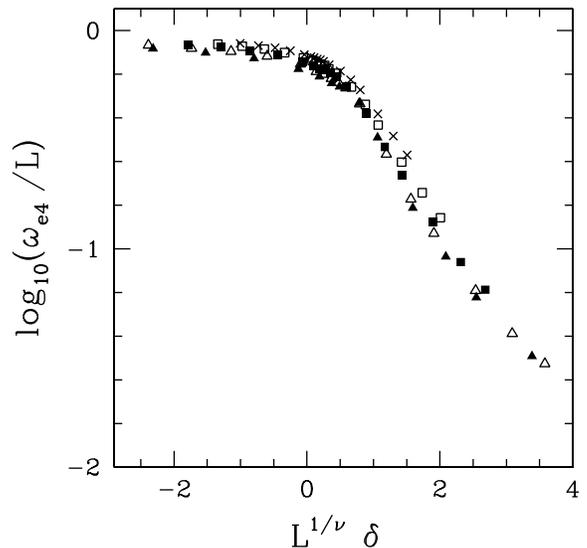}}
\end{picture}
\caption{Finite size scaling collapse of $w_{e4}$
using $\Delta_c=2.52$ and $\nu = 2.4$.
The symbols used for each $L$ are the same as in Figure \ref{fig-w}. 
}
\label{fig-fssw}
\end{figure}

The contribution of small fingerwidths decreases if one calculates
higher moments of the fingerwidth.
We define
\begin{equation}
w_{en} = \root n \of{ \langle w_e ^n \rangle~, } 
\label{w4}
\end{equation}
where $n=1$ gives the mean width, $n=2$ gives the rms width, etc..
We find a steady improvement in finite-size scaling collapses
with increasing $n$.
Figure \ref{fig-fssw} shows that results for $w_{e4}/L$
collapse onto a universal curve at large $L$
with the same $\Delta_c$ and $\nu$ used in Fig. \ref{fig-fssh}.
Best fits for $\nu$ increased consistently from $2.0\pm 0.2$ at $n=1$ to
$2.2 \pm 0.2$ for $n=4$, and the quality of the collapse
showed progressive improvement.

If $w_e$ is proportional to the diverging correlation length,
then it must diverge with the same exponent.
To check this, we examined the slope of plots of $\log_{10} w_{en}$
against $\log_{10} \delta$.
The slopes were indeed consistent with values of $\nu$ from
finite-size scaling,
although the uncertainties were somewhat larger.

\section{Summary and Discussions}
\label{sec:5}

We have studied the zero-temperature
phase diagram for interface growth in the 3D RFIM
with a Gaussian distribution of random fields. 
There is a single transition from self-affine growth below $\Delta_c$
to percolative growth at larger disorder.
The tails of the Gaussian distribution eliminate
the faceted regime observed in previous 3D studies of
bounded field distributions.\cite{ji3d}
This means that the Gaussian RFIM provides a more realistic description
of transitions in systems that do not have an underlying crystalline
lattice, such as random porous media or amorphous magnets.
The phase diagram is also quite different from that for the 2D RFIM
where Gaussian randomness suppresses the self-affine regime.

The critical behavior at the onset of motion in the self-affine
and self-similar regimes was analyzed using finite-size scaling.
The critical exponent $\nu_h$ that describes the diverging
length scale as $H \rightarrow H_c$ was found to be $0.75 \pm 0.02$
for self-affine growth and $0.88 \pm 0.02$ for self-similar growth.
These values are consistent with results for bounded distributions
of disorder.\cite{ji3d}
We also found the same roughness exponent $\alpha = 2/3$
in the self-affine regime (see Fig. \ref{fig-rho}).
These results indicate that changing the form of the distribution
of random fields does not change the universality class of the
self-affine and self-similar growth regimes.

The multi-critical point that separates self-affine and percolative growth
was also analyzed.
We found that the fingerwidth used in previous work does not diverge and
can not be used to determine $\Delta_c$.
Examination of the external interface revealed two lengths that did diverge
at $\Delta_c$:
The overhang size $dh$ diverges as $\Delta$ increases to $\Delta_c$ in
the self-affine regime, and the external fingerwidth $w_e$ diverges as
$\Delta$ decreases to $\Delta_c$ in the percolative regime.
Finite-size scaling collapses of these and other quantities gave
consistent values for $\Delta_c = 2.52 \pm 0.03$ and $\nu=2.4\pm 0.4$.
The error bars on these quantities are estimates of systematic uncertainties
due to corrections to scaling.

The value of $\nu$ determined previously\cite{ji3d} for a bounded distribution
random fields, $\nu=3.0\pm 0.5$, is consistent with our result.
However, this value was determined from the fingerwidth and is not reliable.
Future work is needed to determine whether bounded and unbounded
distributions are in the same universality class.

Perkovi\'c {\em et al} \cite{sethna1,sethna2} have determined the critical
behavior for the 3D Gaussian RFIM using a growth algorithm that appears
to be in a different universality class.
They analyzed the integrated avalanche size distribution
occurring in one branch of a hysteresis loop 
($H$ increasing from $-\infty$ to $\infty$).
From the divergence as $\Delta$ decreased to $\Delta_c$ 
they found numerical values for 
the critical disorder, $\Delta_c = 2.16\pm 0.03$, and 
correlation length exponent, $\nu=1.43 \pm 0.18$.

The discrepancy between our results and those of Perkovi\'c {\em et al.}
seems to result from a crucial difference in our growth algorithms.
They allowed any spin-flip that lowered the energy,
while we only allowed spins on the interface to flip.
The exchange coupling between neighbors dominates in the low disorder limit,
and spins are very unlikely to flip unless they are on the interface.
Nowak {et al.}\cite{nowak1,nowak2} found that the difference between
the two algorithms was negligible in simulations
of low disorder growth with uniform disorder.
The two algorithms should also yield the same percolating cluster in
the high disorder limit, where interactions between neighbors become
irrelevant.
The problem maps onto ordinary percolation, and the order in which spins
are flipped becomes irrelevant.
Near $\Delta_c$ these arguments break down, and the algorithms may give
different results.
At intermediate disorder, the exchange coupling is weak enough to
allow clusters to flip ahead of the interface, and correlations are
important enough that these flipped clusters can aid the advance
of the approaching interface.
One expects that both $H_c$ and $\Delta_c$ will be lowered by the
advance clusters in Perkovi\'c {\em et al.}'s model,
and this is consistent with the numerical results.

\acknowledgements
This work was partially supported by CNPq, CAPES, FAPERJ and FUJB (Brazil),
by National Science Foundation Grant DMR 9634131,
and by Intel Corporation through the donation of workstations that
were used for our simulations.
We thank G. Magnusson for assistance in implementing the growth algorithm,
and R. Paredo and C. S. Nolle for useful conversations.


\begin{thebibliography}{99}
\bibitem{foot1}
Note that the value of $P_c$ depends on the ratio between the height
and width of the cell, but $H_c$ does not (Ref. \cite{nollethesis}).
\end{thebibliography}
\end{document}